\documentclass[aps,prl,reprint]{revtex4-2}

\usepackage{graphicx}
\makeatletter
\let\saved@includegraphics\includegraphics
\AtBeginDocument{\let\includegraphics\saved@includegraphics}
\renewenvironment*{figure}{\@float{figure}}{\end@float}
\makeatother

\usepackage{dsfont}
\usepackage{color}
\usepackage{dcolumn}
\usepackage{bm}
\usepackage{braket}
\usepackage{subfigure}
\usepackage{amsmath,amsfonts,amsthm,amssymb}
\usepackage{times}
\usepackage{lineno} 
\usepackage{graphicx}\graphicspath{{./figures/}}
\usepackage{xcolor}
\usepackage[bookmarks=true, colorlinks=black]{hyperref}
\usepackage{todonotes}

\newcommand{\fig}[1]{Fig.~\ref{#1}}

\newcommand{\Yb}{$^{171}$Yb$^+$}

\begin{document}
\title{Measuring a single atom's position with extreme sub-wavelength resolution and force measurements in the yoctonewton range}
\author{P. H. Huber}
\affiliation{Department Physik, Naturwissenschaftlich-Technische Fakult\"at, Universit\"{a}t Siegen, Walter-Flex-Str. 3, 57072 Siegen, Germany}
\author{P. Barthel}
\affiliation{Department Physik, Naturwissenschaftlich-Technische Fakult\"at, Universit\"{a}t Siegen, Walter-Flex-Str. 3, 57072 Siegen, Germany}
\author{Th. Sriarunothai}
\affiliation{Department Physik, Naturwissenschaftlich-Technische Fakult\"at, Universit\"{a}t Siegen, Walter-Flex-Str. 3, 57072 Siegen, Germany}\author{G. S. Giri}
\affiliation{Department Physik, Naturwissenschaftlich-Technische Fakult\"at, Universit\"{a}t Siegen, Walter-Flex-Str. 3, 57072 Siegen, Germany}
\author{S. W{\"o}lk}
\affiliation{Department Physik, Naturwissenschaftlich-Technische Fakult\"at, Universit\"{a}t Siegen, Walter-Flex-Str. 3, 57072 Siegen, Germany}
\author{Ch. Wunderlich}
\email{Corresponding author: christof.wunderlich@uni-siegen.de}
\affiliation{Department Physik, Naturwissenschaftlich-Technische Fakult\"at, Universit\"{a}t Siegen, Walter-Flex-Str. 3, 57072 Siegen, Germany}
\date{\today}

\begin{abstract}

The center-of-mass position of a single trapped atomic ion is measured and tracked in time with high precision. Employing a near-resonant radio frequency field of wavelength 2.37 cm and a static magnetic field gradient of 19 T/m, the spatial location of the ion  is determined with 
an unprecedented wavelength-relative resolution of 5 $\times$ 10$^{-9}$, corresponding to an absolute precision of 0.12 nm.
Measurements of an electrostatic force on a single ion demonstrate a sensitivity of 2.2 $\times$ 10$^{-23} ~\text{N}/\sqrt{\text{Hz}}$. The real-time measurement of an atom's position complements the well-established technique of scanning near-field radio frequency transmission microscopy and opens up a novel route to using this method with path breaking spatial and force resolution. 
\end{abstract}

\maketitle

\label{sec:introduction}

Experimental techniques enabling high spatial resolution of atoms, molecules, and larger particles are indispensible tools for investigating microscopic and nanoscopic details of matter, and, therefore are of fundamental interest in different branches of science. Scanning near-field optical microscopy \cite{Pohl1984} has been exploited extensively at wavelengths in the visible regime \cite{Trautman1994} and has been extended to the radiofrequency (RF) regime \cite{Keilmann1996,Kramer1996}, to spatially resolve sub-micron features of matter. The highest wavelength-relative resolution attained to date was reported by Keilmann et al. \cite{Keilmann1996}  as $\Delta x / \lambda < 5 \times 10^{-7}$ using wavelengths up to 20 cm ($2\pi \times 1.5$ GHz). Recently,  electron spins in single nitrogen vacancy defect centers in diamond were addressed selectively \cite{Zhang2017}, and were used for measuring magnetic fields, with a wavelength-relative resolution of $2.8 \times 10^{-6}$ using a wavelength of 10.4 cm (corresponding to $\lambda/c =  2.88$ GHz, with velocity of light $c$) \cite{Jakobi2016}.
Combining scanning force microscopy and magnetic resonance imaging was proposed  in 1991 \cite{Sidles1991} and demonstrated shortly after \cite{Rugar1992}. Todays implementations of this technique achieve a resolution better than 10 nm \cite{Degen2008}.
Trapped atomic ions have been succesfully employed as ultrasensitive probes for magnetic fields \cite{Kotler2011,Baumgart2016,Ruster2017}, electric fields, and forces in the yoctonewton  regime \cite{Knunz2010,Biercuk2010,Maiwald2012,Gloger2015,Blums2018}. 

Here, using a single trapped atomic ion exposed to a static magnetic field gradient and an RF field, we demonstrate a wavelength-relative spatial resolution that is two orders of magnitude higher compared to the best reported result to date \cite{Keilmann1996}.
A transition between hyperfine states of the ion's electronic ground state is (near-)resonantly driven by RF radiation. The final state is then determined by detecting state-selectively scattered resonance fluorescence. Thus, the ion's hyperfine splitting can be measured accurately using RF-optical double resonance spectroscopy  and the magnetic field strength at its location is deduced from this measurement. Since the ion is exposed to a spatially varying magnetic field with known gradient, its position can be inferred from the measured  magnetic field strength. By applying an efficient two-point measurement of the ion's RF resonance, we demonstrate how the ion's position can be tracked in real time with a wavelength-relative resolution of 5 $\times$ 10$^{-9}$. This Letter is organized as follows: First, the experimental apparatus and the measurement method are described. Then the precise determination of the spatial location of a single ion is reported. Finally, the results are summarized. 

\label{sec:experimental-setup}

\begin{figure*}
	\centering
		\includegraphics[width=0.95\textwidth]{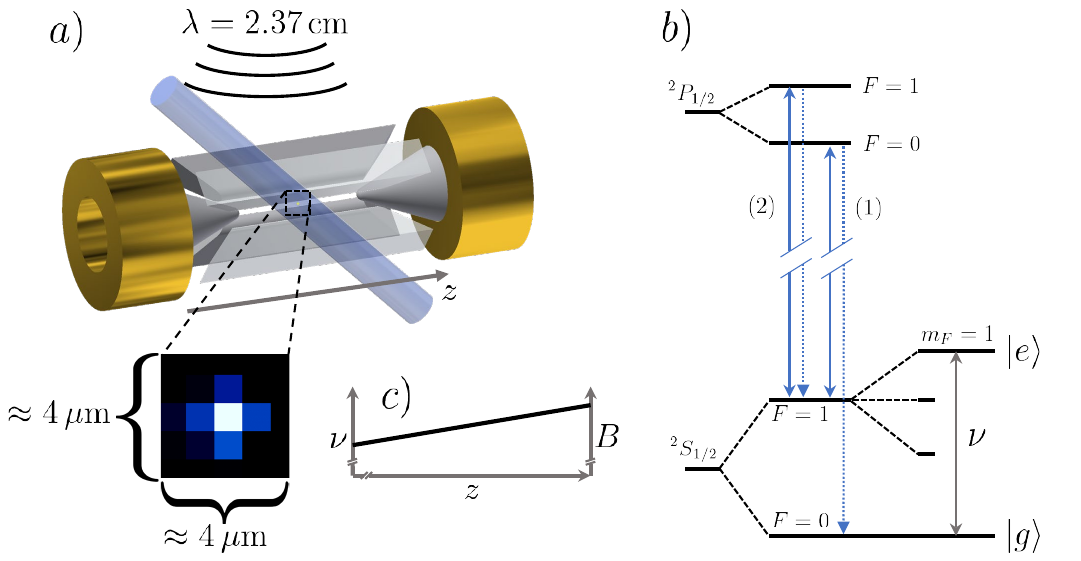}
	\caption{(a) A single \Yb \ ion is confined in a linear Paul trap formed by two pairs of RF electrodes (in light grey, radial direction) and a pair of DC electrodes (in light grey, $z$-direction). A pair of cylindrical permanent magnets mounted on the DC electrodes generates a static magnetic field gradient $\partial_z B= 19.07(2)$ T/m along the $z$-axis. The ion is irradiated by an additional RF field with wavelength 2.37 cm (12.6 GHz) and by laser light. The inset shows a single ion's resonance fluorescence near 369 nm imaged onto an EMCCD camera.
(b) Partial energy level scheme of \Yb (not to scale), showing the ground state hyperfine splitting. The  hyperfine manifold of the ground state $F$=1 level is Zeeman split by an offset magnetic field. The RF resonance $\ket{g} \leftrightarrow \ket{e}$  at angular frequency $\nu$ near of $2\pi \times 12.6$ GHz is magnetic field sensitive to first order. The optical resonances near 369 nm used for initialization of the ion in state $\ket{g}$ (1),  and for Doppler cooling and state selective detection of state $\ket{e}$ (2) are also shown. (c) The magnetic field gradient results in a position dependent frequency $\nu$ of the $\ket{g} \leftrightarrow \ket{e}$ resonance. }
		\label{fig:setup}
	\end{figure*}

{\it Experimental Method.} A single $^{171}$Yb$^{+}$ ion is spatially confined in a macroscopic linear Paul trap~(Fig.\ref{fig:setup}) . The effective harmonic trapping potential is characterized by axial and radial secular trap frequencies of 
$ \omega_z = 2 \pi \times 108.104(8) ~\text{kHz}$ and  $\omega_r =2 \pi \times 534.4(1) ~\text{kHz}$, respectively. The ion is exposed to an offset magnetic field of $|\vec{B}|$ = 442.09(1)~$\mu$T and a static magnetic field gradient of $|\partial_z B|$ = 19.07(2)~T/m applied along the axial direction  \cite{Khromova2012}. The offset field lifts the degeneracy of the hyperfine manifold, while the gradient, in addition, gives rise to  a position dependent energy splitting. 

 The spatial location of the ion is deduced from a measurement  of the hyperfine resonance frequency,  $\nu$ near 12.6 GHz corresponding to the magnetic dipole transition \\ $\ket{^2S_{1/2}\,,F=0\,} \equiv \ket{g} \leftrightarrow \ket{^2S_{1/2}\,,F=1, m_F=+1\,} \equiv \ket{e}$ in the elctronic ground state. Here, $F$ is the quantum number characterizing the total angular momentum and $m_F$ indicates the magnetic quantum number (\fig{fig:setup}). 
 The resonance frequency, $\nu$  is determined by the magnetic field at the location of the ion's center of mass.
 
In order to implement RF-optical double-resonance spectroscopy \cite{Brossel1949}, the ion is first Doppler cooled reaching a mean phonon number $\overline{n} \approx  80$ in the axial trapping potential. 
The ion is then initialized in state $\ket{g}$ by applying laser radiation near 369 nm tuned to the $\ket{^2S_{1/2}\,,F=1\,} \leftrightarrow \ket{^2P_{1/2}\,,F=1}$ resonance. Then, a single RF pulse of fixed duration $\tau$ and with variable  frequency $\nu_{\text{\tiny{RF}}}$ around the resonance $\nu$ of the transition $\ket{g} \leftrightarrow \ket{e}$ is applied, generating a superposition state  $\alpha \ket{g} + \beta \ket{e} $ ($\alpha, \beta \in \mathds{C} $). To measure the RF resonance frequency, the excitation probability $\vert\beta\vert^2$ of state $\ket{e}$ has to be determined after the RF pulse is applied. This is achieved by applying  a  laser field near $369$~nm now resonant with the transition between states $\ket{e}$ and  $\ket{^2P_{1/2},F=0}$. An ion in state $\ket{e}$ scatters resonance fluorescence, an ion in state $\ket{g}$ does not. The ion's resonance fluorescence is collected and imaged onto an electron-multiplying charge coupled (EMCCD) camera. The ion's internal state is projected on $\ket{e}$ or $\ket{g}$ depending on the outcome of this measurement \cite{Woelk2015}. This projective measurement is repeated typically 50 times to estimate $\vert\beta\vert^2$. 

\begin{figure}
	\centering
	\subfigure[]{
		\includegraphics[width=0.45\textwidth]{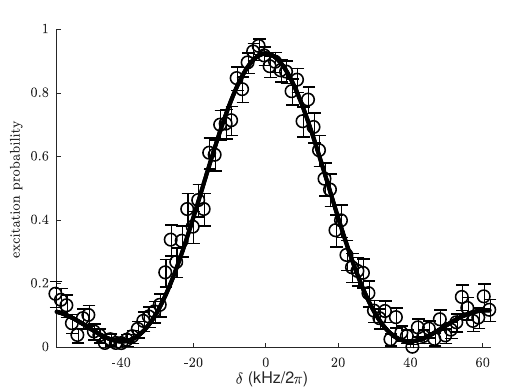}
		\label{fig:rfodr}}
	\subfigure[]{
		\includegraphics[width=0.45\textwidth]{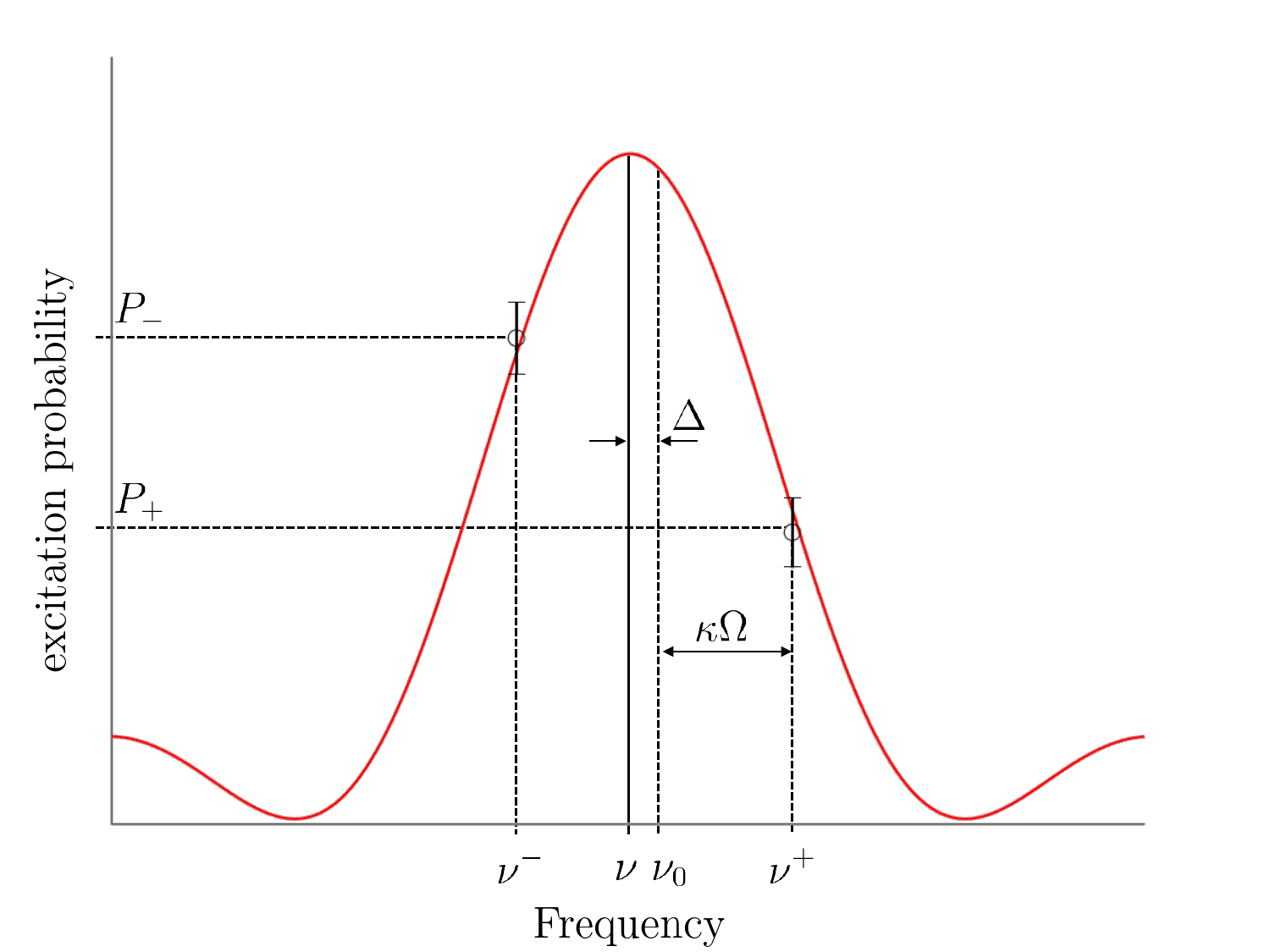}
		\label{fig:resonance}}
	\caption{(a) Exemplary spectrum measured by RF-optical double resonance spectroscopy showing the  $\ket{g} \leftrightarrow \ket{e}$ resonance in the hyperfine manifold of \Yb  ($\nu$ = $\nu_{RF} - 2\pi \times 12648759.8(3)$ kHz). 
The Rabi frequency is $\Omega$ = $2\pi \times 25$ kHz. The measurement is carried out in 80 frequency steps of size $2\pi \times 1.5$ kHz and 100 repetitions for each frequency step. The solid red line represents a fit to the experimental data (Eq.~2 in Supplemental Material (SM)). The error bars represent one standard deviation due to quantum projection noise. (b) Shape of the simulated atomic resonance (Eq.~2 in SM). Here, $\nu_0$ is the initial estimate of the resonance frequency, while $\nu$ is the actual resonance frequency. Two points of the resonance curve at frequencies $\nu^\pm=\nu_0\pm \kappa\Omega$ are measured.  $\Delta=\nu-\nu_0$ is the frequency offset to be determined.  $2\kappa\Omega$, corresponding to the FWHM of the resonance line, is the frequency separation between the two measuring points, symmetrically chosen around $\nu_0$.}
		\label{resonance-spectrum}
	\end{figure}

 The excitation probability $\vert\beta\vert^2$ after an RF pulse with duration $\tau$ and frequency $\nu_{\text{\tiny{RF}}}$  depends on the detuning $\delta=\nu -\nu_{\text{\tiny{RF}}}$ from the resonance frequency. The duration $\tau$ of the RF pulse is chosen such that it corresponds to a $\pi$ pulse inverting the population of the states $ \ket{g}$ and $ \ket{e}$ when $\delta=0$, that is,  $\tau=\tau_\pi=\pi/\Omega_0$ where $\Omega_0$ is the Rabi frequency on resonance.  
An exemplary measurement of the hyperfine resonance $\ket{g} \leftrightarrow \ket{e}$   is shown in \fig{fig:rfodr}. A fit of the experimental data allows for extracting the center frequency $\nu$ of the resonance line and thus the ion's position in the magnetic gradient field. 

A laser beam with beam waist of  $ 140~\mu$m is used to drive the optical resonance  $\ket{^2S_{1/2}\,,F=1} \leftrightarrow \ket{^2P_{1/2}\,,F=1}$ with a natural line width of $2\pi \times 19.6$~MHz (Fig.~\ref{fig:setup}). Imaging an ion's resonance fluorescence onto the EMCCD camera allows for the reconstruction of the ion's center-of-mass position with a spatial resolution of about $35$~nm   \cite{Gloger2015,Blums2018}. Here we demonstrate position measurements with a spatial resolution that  is better by more than two orders of magnitude. 

Lowering the Rabi frequency, by reducing the amplitude of the applied RF field, narrows down the resonance, thus enhancing the precision of determining $\nu$ and consequently of the position measurement. 

Excitation of the ion motion in the trapping potential will change the shape and width of the resonance, since the Rabi frequency depends on the ion's vibrational excitation quantified by quantum number $n$. This is taken account when fitting the resonance line in \fig{fig:rfodr}. 
Varying the mean phonon number $\overline{n}$ of a thermal distribution in the range from 20 to 100 phonons yields a variation of the FWHM of the RF resonance by about 1 \%  making Doppler cooling sufficient for the precision of the position measurement desired in this work.  

When tracking an ion's position in real time, we determine its current resonance frequency $\nu$ from the measurement of the excitation probability $\vert\beta\vert^2$ for only two values of the detuning $\delta$. Such a two-point frequency measurement reduces decisively the total measurement time required for determining the center frequency $\nu$ and is therefore more efficient as compared to the routinely employed technique of RF-optical double resonance spectroscopy \cite{Ludlow2015} where the full resonance line is mapped out (\fig{fig:rfodr}).  

These two measurements are carried out with the applied RF field symmetrically detuned such that $\nu_{\text{\tiny{RF}}}^\pm= \nu_0 \pm \kappa\Omega$. Here, $\nu_0$ is the estimated value for the current actual resonance frequency $\nu$ (except for the first measurement, $\nu_0$  is the result of the preceding determination of $\nu$).   $\kappa=0.8$ is chosen such as to separate the two measurement points by the FWHM  $\approx 1.6\times \Omega$ of the hyperfine resonance as shown in Fig.~\ref{fig:resonance}.

\begin{figure}
	
		\subfigure{
		\includegraphics[width=0.45\textwidth]{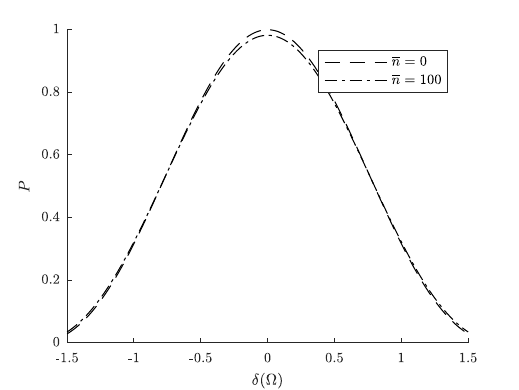}
		\label{fig:FWHM}}
	\subfigure{
		\includegraphics[width=0.45\textwidth]{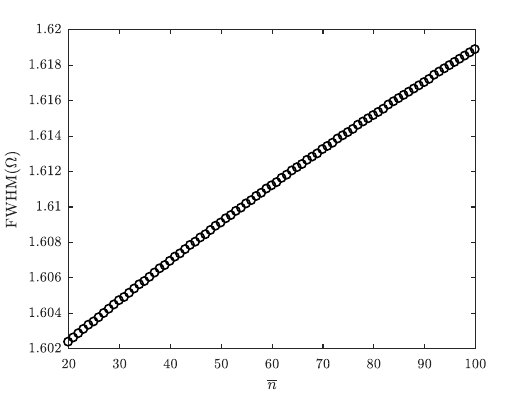}
		\label{fig:FWHM2}											}
\caption{(a) The calculated resonance curve of the transition $\ket{e}\leftrightarrow\ket{g}$ as a function of the detuning $\delta$ for mean phonon numbers $\overline{n} =0 $ and $\overline{n} =100.$ (b)
FWHM of the ion's resonance as a function of the motional thermal excitation $\overline{n}$ in the harmonic trapping potential. The small broadening of the resonance when the ion is motionally excited dispenses with the need for ground state cooling and makes Doppler cooling sufficient, thus allowing for short measurement times. }
\label{fig:fwhm}
\end{figure}

We define {$\Delta=\nu-\nu_0$} to be the initially unknown offset of  $\nu_0$ from the actual resonance frequency $\nu$.  The probabilities for detecting the ion in state $\ket{e}$ when driving the  $\ket{g} \leftrightarrow \ket{e}$ resonance at frequency $\nu_{\text{\tiny{RF}}}^\pm$ are  $P_\pm=P(\overline{n},\nu_{\text{\tiny{RF}}}^\pm)$. The function  
\begin{linenomath}
\begin{equation}
g(\Delta)=\frac{P_+ - P_-}{P_+ +P_-} 
\label{eq:ppmpm}
\end{equation}
\end{linenomath}
maps the measured probabilities to the frequency offset $\Delta$, but is only injective as long as $\nu_{\text{\tiny{RF}}}^\pm$ does not cross an extremal point of the resonance curve. Therefore, $\Delta$ must fulfill $-(1-\kappa)\Omega \leq \Delta \leq (1-\kappa)\Omega $. In this range, $g$ is monotonously increasing with $\Delta$. The calculation of $g(\Delta)$ therefore allows for extracting $\Delta$ by searching for the closest match of the measured values of $g_{\text{meas}}$ and $g_{\text{calc}}(\Delta)$.
The standard deviation of determining $\Delta$ (equal to the standard deviation of the ion's resonance frequency) in the measurements reported here is $\sigma_\nu/\Omega\approx 0.05$ (SM).

With a spatially constant magnetic field gradient $\partial_z B$ \cite{Piltz2016}, a shift in position can be deduced from a change in resonance frequency as
\begin{linenomath}
\begin{equation}
\Delta z = \frac{\Delta B(\Delta)}{ \partial_{z} B}.
\label{eq:position}
\end{equation}
\end{linenomath}
In the present setup,  $\Delta z =1 \text{ nm}$ corresponds to a change in resonance frequency $\Delta \nu \approx 2 \pi \times 266~\text{Hz}$. Repeated measurements of the resonance frequency, employing the method described above, enable the experimenter to follow changes of an ion's position by using the last measured frequency as a reference for the succeeding measurement. 

\label{sec:results}

\begin{figure}
\centering	
\includegraphics[width=0.45\textwidth]{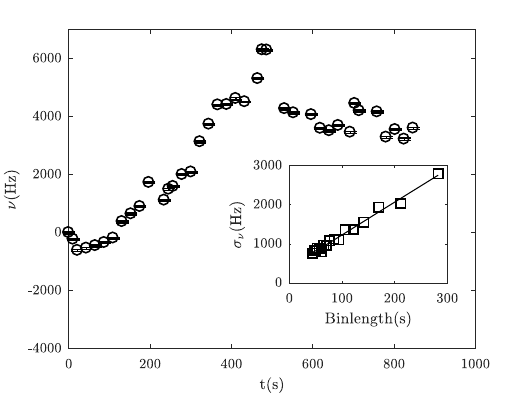}
\caption{The ion's resonance frequency as a function of time without applying a voltage difference, $\Delta U$ along the $z$-direction. This drift is induced by a varying ambient magnetic field and/or uncontrolled movement of the ion in the magnetic field gradient. The inset shows a fit of the Allan variance during the experiment indicating a drift rate of $2\pi \times 8.2$~Hz/s. This uncontrolled drift limits the minimal spatial resolution.}
\label{fig:drift}
\end{figure}

First, we measure the drift of an ion's resonance frequency as a function of time (Fig. \ref{fig:drift}), without applying an additional static electric field that would deterministically shift the ion's position. The observed drift could be caused by changes of  an (uncontrolled) ambient electric field by shifting the ion's position in the magnetic gradient field, and/or by an (uncontrolled)  change of the ambient magnetic field at the ion's position.    
The drift rate of $2\pi \times 8.2$~Hz/s during this experiment is calculated using a fit to the Allan variance (Inset of Fig. \ref{fig:drift}) based on the  measured frequencies shown in Fig. \ref{fig:drift}. The drift during the total time of 2 s needed to measure ion's resonance frequency then amounts to $2\pi \times 16.4~$Hz which translates into an uncertainty in determining the ion's position of $0.12$~nm.  This limits the spatial resolution of the current setup.
Applying shielding for electric and magnetic fields \cite{Ruster2016}, which is not present in the current experimental setup, will further increase the precision in determining an ion's position.

\begin{figure}
	
		\subfigure{
		\includegraphics[width=0.45\textwidth]{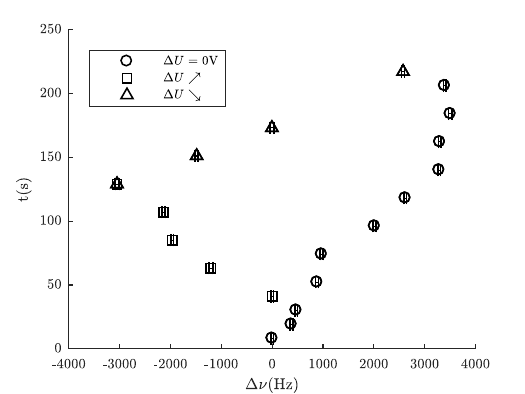}
		\label{fig:tz}}
	\subfigure{
		\includegraphics[width=0.45\textwidth]{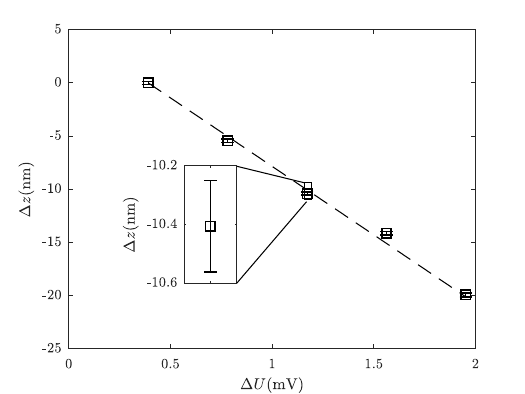}
		\label{fig:zU}}
		
	\caption{(a) Resonance frequency of a single ion. Circles indicate measurements tracking the frequency of a drifting single ion (Fig. \ref{fig:drift}).  For the measurements indicated by squares (triangles), an increasing (decreasing) voltage  $\Delta U \neq 0$ is applied to the DC electrodes purposely shifting the ion's position along the $z$-direction. (b) Change in position of a single ion as a function of the applied voltage $\Delta U$ after subtracting the drift. The dashed line shows the calculated displacement of the ion along the $z$-direction in agreement with the measured data. The errorbars represent one standard deviation. The mean position uncertainty of $\overline{\sigma_z} =$ 0.12 nm is calculated as the arithmetic mean of the standard errors for these measurements.}
		\label{fig:position}
	\end{figure}

To demonstrate the capability of tracking an ion's position, an imbalanced voltage, $\Delta U$ of varying magnitude is applied to the DC trapping electrodes shifting the ion's equilibrium position along the z-axis (Fig. \ref{fig:setup}). Measurements of the ion's resonance frequency, $\nu$ while applying  $\Delta U \neq 0$ V are alternated with measurements where  $\Delta U=0$ V  to correct the electric field-induced change in position for an unwanted drift. 
Fig.~\ref{fig:tz} shows the change of the ion's resonance as a function of time.  The measurements with $\Delta U =0$ in Fig.~\ref{fig:tz} indicate the unintended drift due to magnetic field drifts and possible voltage drifts during the experiment (the data points shown here are the same as in Fig.~\ref{fig:drift}, corresponding to the time between $t=235$ s and $t=433$ s.) The change of the resonance frequency  during the measurements with applied voltage, $\Delta U \neq 0$ is interpolated between consecutive measurements with $\Delta U =0$ and subtracted from the absolute change in frequency. The resulting change in the ion's position as a function of $\Delta U$ is shown in Fig.~\ref{fig:zU}. In addition, Fig.~\ref{fig:zU} shows the change in the ion's position calculated as a function of  $\Delta U$ (dashed line) that matches well the measurements of the voltage-induced drift. 

The standard deviation in determining the center of the resonance line in each measurement shown in Fig. \ref{fig:tz}  translates into a standard deviation $\sigma_z$ of the ion's position (SM, Eq.~1 and error propagation) as shown in Fig. ~\ref{fig:zU}. The mean standard deviation of all measurements of the ion's position is  $\overline{\sigma_z} =$ 0.12 nm, and therefore yields a sub-wavelength resolution of $\overline{\sigma_z} / \lambda \leq 5\times 10^{-9}$.

From this position measurement, having characterized the trapping potential independently, a force displacing the ion can be derived. 
For the effective harmonic trapping potential used here, the constant of proportionality between displacement and force, $k_z$ is given by  
$ k_z=m_{\text{Yb}} \omega_z^2=1.3\times 10^{-13} ~\text{N/m}$. 
This gives a force resolution of $\sigma_F = k_z \sigma_z = 1.5 \times 10^{-23} ~\text{N}$. Thus, for a measurement duration of 2~s we achieve a sensitivity of this static force measurement of $2.2\times 10^{-23} ~\text{N}/\sqrt{\text{Hz}}$. 
This force resolution would allow, for example, to detect a single elementary charge at a distance of $6~\text{mm}$. 

Current limitations in the measurement time and the resolution of the position measurement can be overcome by continuous sympathetic cooling of the ion to be tracked by a second ion species. This will remove the necessity for Doppler cooling between measurements. Application of magnetic shielding will remove the necessity for maintaining a fixed phase relation of the measurement cycles with respect to the power line, thus further shortening the measurement time. A reduction of the uncontrolled ion drift by shielding against electric and magnetic fields will make it possible to carry out measurements at lower Rabi frequencies. We estimate that, using the method introduced here, the precision of position and force measurements could be enhanced further by several orders of magnitude (see SM). 

We believe that the technique presented in this Letter can be adapted to other atomic and molecular ion systems with modest experimental effort. For example, this technique could be applied in the context of precision frequency metrology in order to detect, characterize and compensate for small shifts in the spatial position of single atomic \cite{Schmidt2005} and molecular \cite{Wolf2016, Chou2017, Sinhal2020} ions. One may speculate about applying this technique to the detection of charges in particle physics. \cite{Carney2021, Budker2021}. 

While direct application of the demonstrated method in microscopy is beyond the capability of our current trap design, possible future devices can employ novel trap designs, for example a stylus trap \cite{Maiwald2009}, as a scanning probe with wavelength-relative spatial resolution orders of magnitude better than any state-of-the-art device known to us.

\begin{acknowledgements}
The authors acknowledge funding from Deutsche Forschungsgemeinschaft and from the Bundesministerium f{\"u}r Bildung und Forschung (FK 16KIS0128). G.~S.~G. was supported by the European Commission's Horizon 2020 research and innovation program under the Marie Sk\l{}odowska-Curie grant agreement number 657261.
\end{acknowledgements}

\bibliographystyle{apsrev4-2}
\bibliography{paper}

\end{document}


\title{Supplemental Material: Measuring a single atom's position with extreme sub-wavelength resolution and force measurements in the yoctonewton range}
\author{P. H. Huber}
\affiliation{Department Physik, Naturwissenschaftlich-Technische Fakult\"at, Universit\"{a}t Siegen, Walter-Flex-Str. 3, 57072 Siegen, Germany}
\author{P. Barthel}
\affiliation{Department Physik, Naturwissenschaftlich-Technische Fakult\"at, Universit\"{a}t Siegen, Walter-Flex-Str. 3, 57072 Siegen, Germany}
\author{Th. Sriarunothai}
\affiliation{Department Physik, Naturwissenschaftlich-Technische Fakult\"at, Universit\"{a}t Siegen, Walter-Flex-Str. 3, 57072 Siegen, Germany}\author{G. S. Giri}
\affiliation{Department Physik, Naturwissenschaftlich-Technische Fakult\"at, Universit\"{a}t Siegen, Walter-Flex-Str. 3, 57072 Siegen, Germany}
\author{S. W{\"o}lk}
\affiliation{Department Physik, Naturwissenschaftlich-Technische Fakult\"at, Universit\"{a}t Siegen, Walter-Flex-Str. 3, 57072 Siegen, Germany}
\author{Ch. Wunderlich}
\email{Corresponding author: christof.wunderlich@uni-siegen.de}
\affiliation{Department Physik, Naturwissenschaftlich-Technische Fakult\"at, Universit\"{a}t Siegen, Walter-Flex-Str. 3, 57072 Siegen, Germany}
\date{\today}
\maketitle

\section{Methods}
\label{sec:methods}

The  dependence of the ion's hyperfine resonance frequency on the magnetic field is  described by the Breit-Rabi formula \cite{Breit1931}
\small
\begin{linenomath}
\begin{align}
h \nu = g_k\mu_K B + \frac{A}{2}\sqrt{1+XB+X^2B^2} + \frac{A}{2}\sqrt{1+X^2B^2},
\label{eq:breit-rabi}
\end{align}
\end{linenomath}
\normalsize
where $g_k$ = 0.9837 is the Land\'{e} g-factor of the nucleus, $\mu_K$ is the nuclear magneton, $A = 2\pi \times 12642812118.471(9)$ Hz is the hyperfine constant of $^{171}$Yb$^{+}$\cite{Pendrill1994}, $g_j$ is the Land\'{e} g-factor of the electron, $\mu_B$ is the Bohr magneton, and $X= (g_j \mu_B - g_k \mu_k)/A$. Thus, the absolute value of the magnetic field $B$ at the ion's position can be deduced from measuring the atomic resonance frequency $\nu$. Utilizing the magnetic field gradient $\partial_zB$,  a measurement of the resonance frequency can be mapped to a position in space. 
The magnitude of the magnetic field gradient has been determined to be $\partial_z B =19.07(2)\text{\,T/m}$ by measuring the resonance frequencies of a string of eight trapped ions calculating the magnetic field from the measured frequencies and their calculated positions by numerically minimizing the total energy. 

Doppler cooling is achieved by applying laser light near $369$~nm, driving the resonance between the states $ \ket{^2S_{1/2}\,,F=1}$ and  $\ket{^2P_{1/2}\,,F=0}$, and RF radiation driving the resonance $\ket{^2S_{1/2}\,,F=0\,} \leftrightarrow \ket{^2S_{1/2}\,,F=0\,,m_F=+1}$, thus avoiding optical pumping into the state  $\ket{^2S_{1/2}\,,F=0}$ \cite{Sriarunothai2018}. Optical pumping into the long-lived meta-stable state $\ket{^2D_{3/2}}$ is prevented using laser light near $935$~nm and pumping  into the state $\ket{^2F_{7/2}}$ is prevented using laser light near $638$~nm, respectively. 

The shape of the atomic resonance $\ket{g} - \ket{e}$ depends on the Rabi frequency $\Omega_n$, and we have  
\begin{equation}
P_n(\delta / \Omega_n)=\frac{1}{1+\left( \frac{\delta}{\Omega_n}\right)^2} \sin^2\left(\sqrt{1+\left(\frac{\delta}{\Omega_n}\right)^2} \frac{\tau_\pi}{2}\right)
\label{eq:norm}
\end{equation}
with excitation probability $P_n$. $\Omega_n$ is the Rabi frequency of the transition when the ion populates the $n^\text{th}$ vibrational state of the harmonic trapping potential.
A thermal distribution of vibrational states is taken into account when fitting resonance lines as 
\begin{linenomath}
\begin{equation}
P(\overline{n},\delta)=\sum_{n=0}^{10\overline{n}}{\frac{1}{\overline{n}+1}}\left( \frac{\overline{n}}{\overline{n}+1} \right )^nP_n(\delta).
\label{eq:thermalprob}
\end{equation}
\end{linenomath}
The width of the observed transition as a function of the thermal excitation $\overline{n}$ is shown in Fig. 3(a) and 3(b) of the main text. Varying the mean phonon number in the range from 20 to 100 phonons yields a variation of the FWHM of the RF resonance from $1.602 \Omega$ to $1.62 \Omega $. The experiments presented here were obtained using Doppler cooling to mean vibrational excitation, $\overline{n} \approx 80$ \cite{PiltzPhD,Sriarunothai2018}.

\begin{figure}
\centering
\includegraphics{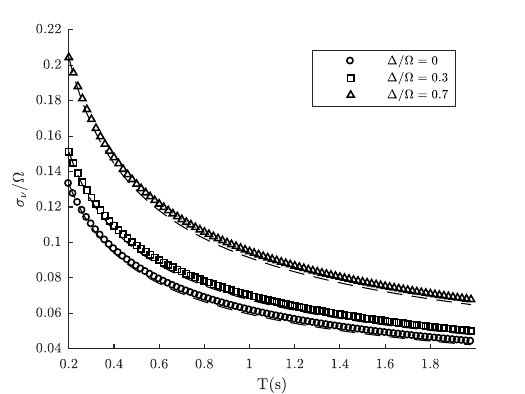}
\caption{Calculated standard deviation in units of the Rabi frequency $\Omega$ when determining the ion's resonance frequency as a function of measurement time $T$ for a given offset $\Delta$. Circles indicate the case $\Delta=0$, that is  the initial guess of the resonance frequency is exact. Squares and triangles indicate that the measured resonance frequency is differing from the expectation by $\Delta/\Omega=0.3$ or $0.7$ respectively. The standard deviation scales with the total measurement time as $T^{-1/2}$. }
\label{fig:deltaoft}
\end{figure}

The standard deviation of the frequency and position measurements reported here is a function of the offset $\Delta$ as well as of the measurement duration. Fig. \ref{fig:deltaoft} shows the simulated uncertainty of a frequency measurement based on quantum projection noise and error propagation as a function of the total measurement time $T$. Here, $T$  includes idle periods between repetitions of individual measurements of the ion's hyperfine state. 

A measurement of the resonance frequency  $\nu$ as reported in this work consists of $50$ repetitions for each of the two detunings around the assumed frequency $\nu_0$ resulting in 100 repetitions per measurement in total.  The repetition rate of $50$~Hz is fixed by the fact that each individual measurement is carried out  with a fixed phase with respect to the $50$~Hz  power line to eliminate effects of the magnetic field it generates. For $T=2$~s, the repetition rate allows for measuring the resonance frequency with a standard deviation $\sigma_\nu/\Omega\approx 0.05$. The scaling with time as $T^{-1/2}$ is the same as the fundamental limit for optical localization accuracy \cite{Thompson2002,Ober2004} in microscopy.

Improvements of the current experimental setup will allow for further decisive improvements in the precision of localizing a trapped ion and of measuring small forces. First, shielding the measurement volume against uncontrolled variable electric and magnetic fields should suppress drifts of the measured resonance frequency $\nu$ such that they no longer limit the precision of these measurements. Second,  the effective measurement time can be extended by using a second ion species as a refrigerator ion, thus eliminating the need to interrupt measurement cycles for cooling.  These measures would allow for reducing the Rabi frequency, say to $\Omega\approx 2\pi \times 30 \text{\,Hz}$ giving a frequency resolution, $\sigma_\nu\approx 1.5\text{\,Hz}$ and  a position resolution of $6\times 10^{-12}\text{\,m}$ in  $2$\,s measurement time. Stabilization of the trapping potential would allow to weaken the axial confinement of the probe ion. Assuming a reduction of the force constant to $k_z\approx  4\times 10^{-14} \text{\,N}/\text{m}$ a  force sensitivity of $3.2\times 10^{-25}\text{\,N}/\sqrt{\text{Hz}}$ would be achieved improving the measurement results presented in this paper by about 2 orders of magnitude. Increasing the magnetic field gradient would make further improvements possible.

\bibliographystyle{apsrev4-2}
\bibliography{paper}